\documentclass[a4paper]{article}
\pdfoutput=1
\usepackage{INTERSPEECH2020}
\usepackage{xurl}

\usepackage{hyperref}
\hypersetup{colorlinks,allcolors=black}

\usepackage{multirow}
\usepackage{diagbox}
\usepackage{float}

\usepackage{booktabs}
\usepackage{color}
\usepackage{url}
\usepackage{subfig}
\usepackage{enumitem}

\title{Stereo InSE-NET: Stereo Audio Quality Predictor Transfer Learned from Mono InSE-NET}
\name{Arijit Biswas$^1$, Guanxin Jiang$^{1}$}
\address{
  $^1$Dolby Germany GmbH, Nürnberg, Germany}
\email{arijit.biswas@dolby.com, guanxin.jiang@dolby.com}

\begin{document}

\maketitle
\begin{abstract}
Automatic coded audio quality predictors are typically designed for evaluating single channels without considering any spatial aspects. With InSE-NET~\cite{InSE-NET}, we demonstrated mimicking a state-of-the-art coded audio quality metric (ViSQOL-v3~\cite{v3}) with deep neural networks (DNN) and subsequently improving it – completely with programmatically generated data. In this study, we take steps towards building a DNN-based coded stereo audio quality predictor and we propose an extension of the InSE-NET for handling stereo signals. The design considers stereo/spatial aspects by conditioning the model with left, right, mid, and side channels; and we name our model Stereo InSE-NET. By transferring selected weights from the pre-trained mono InSE-NET and retraining with both real and synthetically augmented listening tests, we demonstrate a significant improvement of $12\%$ and $6\%$ of Pearson's and Spearman's Rank correlation coefficient, respectively, over the latest ViSQOL-v3~\cite{ViSQOLgithub}.
\end{abstract}


\section{Introduction}
\label{section:Introduction}

Multimedia transmission now dominates the internet. While video transmission and storage indisputably command more resources, audio is also a resource hog, and like video, is important to consumers. Modern media consumers are increasingly savvy about audio technology and expect a high quality of experience (QoE) when listening using increasingly high-resolution and high-fidelity systems, whether they be on mobile devices or in their living rooms. Therefore, sound-quality evaluation tests are critical since they provide the necessary user feedback that drives improvements in these technologies. However, subjective listening tests demand a large amount of time and effort. Thus, there is a significant impetus to develop and deploy efficient and accurate audio quality assessment models that can be used to monitor and control end-user QoE. Furthermore, since bandwidth-hungry spatial audio is becoming more pervasive~\cite{NetflixTechBlog}, perceptual audio rate control has become more important and should be a significant factor in QoE optimization, especially when the available bandwidth becomes limited. Many lossy speech and audio codecs are not able to exactly preserve the positions of the various sources within the auditory scene or they may even discard the spatial information altogether if the available data rate is too small~\cite{audio_codec_artifacts},\cite{AESartifacts}. Furthermore, the assessment of binaural audio quality is receiving increasing interest since spatial features in the latest codecs are of increasing importance. Any decrease in spatial quality results in a significant decrease in presence and immersion. These issues greatly motivate us to focus on and develop a consistent and reliable stereo audio quality metric. We are interested in assessing spatial quality in a way that ensures how accurately the sources are positioned as compared to the unencoded reference as well as the effects of audio fidelity degradation due to audio coding.

Quality metrics for audio coding have been primarily designed to evaluate mono audio codecs, the earliest being PEAQ~\cite{PEAQstandard},\cite{peaq}. The fundamental principle of PEAQ is the calculation of so-called model output variables (MOVs) of a monaural hearing model, comparing these MOVs of the reference signal and the degraded signal and feeding these differences into a shallow neural network (NN) that is trained based on the known results of numerous listening tests. PEAQ does offer the possibility to evaluate stereo signals. However, for stereo, two monaural hearing models are used in parallel~\cite{PEAQstandard}, i.e., no inter-channel cues or spatial aspects are considered. The exception is the basic version of PEAQ, where, with the 2 (out of 11) MOVs, a sort of “worse ear”-approach~\cite{PEAQstandard} is used to look for perceivable distortions. Following PEAQ, POLQA~\cite{polqa} and ViSQOL~\cite{Hines20156:ViSQOLAudio01},\cite{7940042},\cite{v3} were developed. POLQA was primarily developed for speech and does not include any spatial aspects in its design. In ViSQOL, left, right, mid, and side channels were evaluated and studied, but ultimately, the final design considered the mid-channel only~\cite{7940042},\cite{ViSQOLgithub}. Thus, none of the popular and standardized audio quality metrics consider spatial aspects in their design. Since the fidelity of the spatial information is not explicitly considered, this may lead to faulty quality estimates~\cite{peaq-mc}. 

Several research papers~\cite{peaq-mc},\cite{peaq-binaural},\cite{seo2013perceptual} have focused on learning a spatial audio quality metric from known binaural auditory cues~\cite{Blauert} like Interaural Level Differences (ILD), Interaural Time Differences (ITD) and Cross- Correlation (IACC) between signals entering the left and the right ear. Although early work to extend PEAQ to spatial audio was published~\cite{peaq-mc} it did not yield a recommendation. Schäfer et al.~\cite{peaq-binaural} presented an extension to PEAQ, that utilized MOVs from a binaural hearing model which is fed to a NN together with the result of PEAQ. In a similar spirit,~\cite{seo2013perceptual} extended PEAQ by a binaural auditory model, allowing predicting the audio quality of multichannel audio codecs.  Delgado and Herre~\cite{Fhg_DirecLoudMap} proposed utilizing distortion between directional loudness maps of reference and coded audio as an indicator for spatial distortion, and in addition, utilized the objective scores predicted by PEAQ and POLQA to predict the overall objective quality.  In their evaluation they specifically excluded speech signals. All the mentioned approaches can be considered as an add-on to the standard PEAQ (and POLQA).   In AMBIQUAL \cite{ambiqual}, researchers presented a full-reference objective audio quality metric designed to estimate the robustness of ambisonic audio compression in terms of perceived quality and localization accuracy. The metric is built on the same similarity measure as used in ViSQOL. However, unlike ViSQOL, a spectrogram of phase angles (rather than magnitudes) is used for signal similarity comparisons.

A recent more powerful alternative is provided using deep-learning-based concepts, which have offered solutions that are accurate, rapidly re-trainable, and easily expandable in many speech and audio-related tasks. In~\cite{smaq_netflix}, the authors leveraged the most important features computed by the perceptual frontend from several audio quality metrics. They trained a multi-task student model using unlabeled data and labels from multiple audio quality metrics; and showed improved generality of the student model in predicting the quality score. Their research follows a similar spirit to our previous InSE-NET~\cite{InSE-NET} research. However, neither of the models was designed for stereo. There exists a deep learning-based perceptual spatial audio quality metric~\cite{SAQAM} that evaluates the similarity of binaural presentations in terms of both localization and sound fidelity degradations between any pair of binaural signals. However, it does not predict an easy-to-interpret (e.g., MUSHRA) quality score. It can only predict (deep feature) distances from the quality-predictive features delivered by a DNN. Furthermore, the model is trained for speech signals at 16~kHz (information about the sampling rate is inferred from their previous paper~\cite{dplm}).

With InSE-NET~\cite{InSE-NET} we demonstrated mimicking a state-of-the-art coded audio quality metric with deep neural networks (DNN) and subsequently improving it – completely with programmatically generated data. We showed that with synthetic data augmentation, one can steer the model to predict accurately. Acknowledging that training with listening tests should further improve the prediction accuracy, in this contribution we: (1) present an extension for handling stereo signals, (2) transfer selected weights from the pre-trained InSE-NET and retrain with listening tests. The design considers stereo/spatial aspects, and we name our model Stereo InSE-NET. We are not aware of any other deep learning-based coded stereo audio quality prediction model handling general audio signals at 48~kHz. Furthermore, we directly predict the subjective quality scores of coded audio signals on a 0-100 MUSHRA~\cite{MUSHRA} quality scale; a well-established quality scale for assessing the quality of audio codecs. Our goal is to demonstrate consistent and reliable coded audio quality prediction, both for waveform preserving codecs (e.g., AAC) and for non-waveform preserving codecs that include parametric bandwidth extension tools (e.g., HE-AAC v1) and parametric stereo coding tools (e.g., HE-AAC v2). In such a use-case of predicting the overall audio coding quality, the interaction of spatial cue distortions along with monaural/timbral distortions (especially in non-waveform-preserving cases) presents a challenging scenario.

The most obvious application of our work is a stand-alone coded audio quality metric, which is useful e.g., for validating encoder improvements with new tunings. Furthermore, since it is based on a deep network, our proposed model can be developed into a learnable comparable loss function for coded audio enhancement~\cite{DCAE}.

The rest of the paper is organized as follows: In Section 2, the data used for model training and evaluation are introduced. Section 3 depicts the details of our proposed Stereo InSE-NET model. The related experimental results and analysis are given in Section 4, and finally, the conclusion is drawn in Section 5.

\section{Datasets}
\label{section:Datasets}

\subsection{Training set}
\label{subsection:trainingset}

\begin{figure}[t]
\begin{center}
  \includegraphics[width=1.1\linewidth]{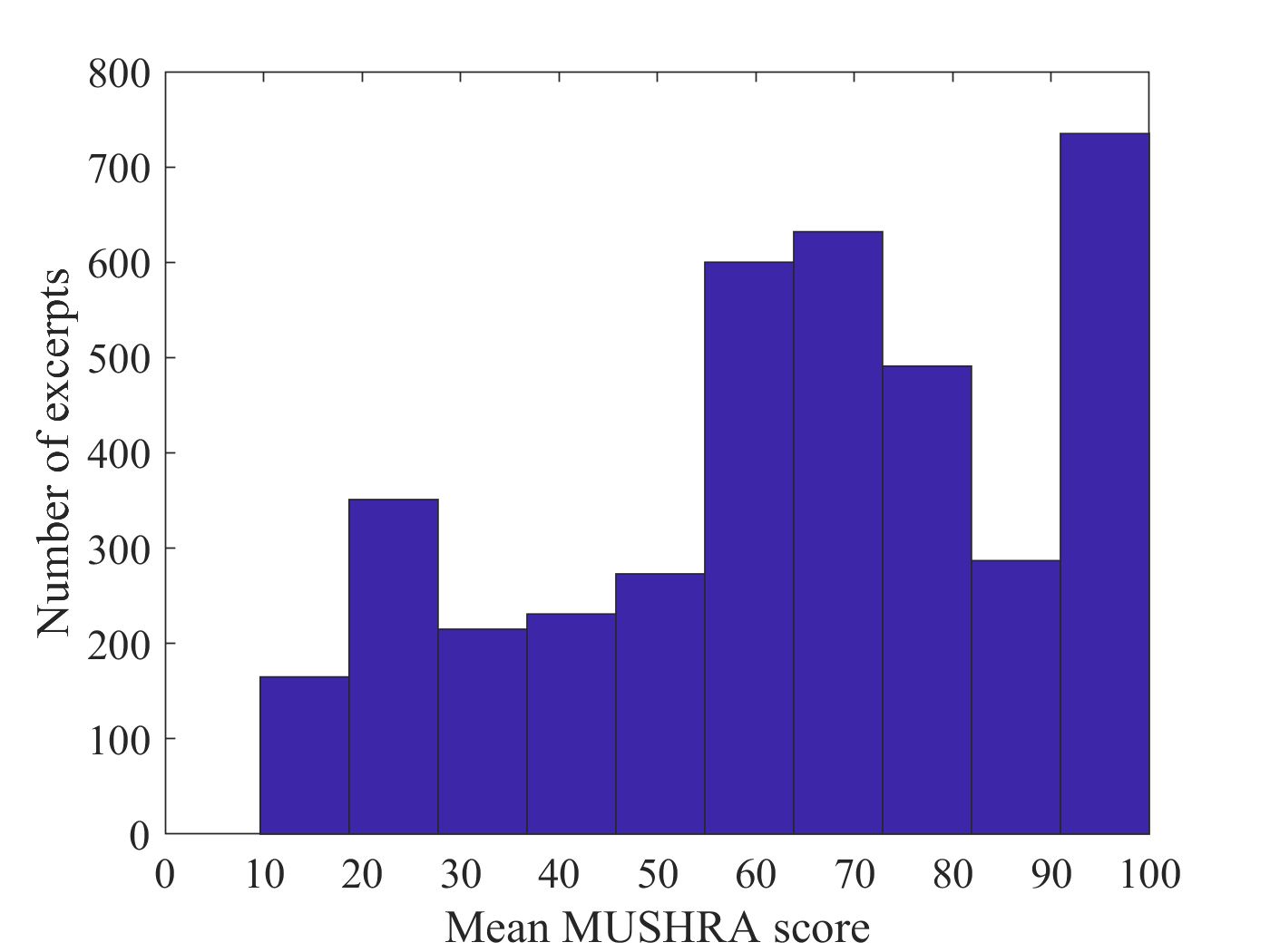} 
  \caption{Histogram of mean MUSHRA scores from the training database.}
  \label{fig:mushraDistribution}
  \end{center}
\end{figure}

When we started this study, we were aware that fine-tuning the mono InSE-NET with listening tests should further improve prediction accuracy. However, the time and expense of obtaining the “gold standard” human judgments limit the availability of such data. For this study, we consciously decided to make the best use of existing listening test data and pre-trained models. We realized we had significantly more stereo listening tests with which to train the model than mono listening tests. Therefore, we decided to enhance and extend the mono InSE-NET to Stereo InSE-NET by training with stereo listening tests.

We used our internal corpus of Multiple Stimuli with Hidden Reference and Anchor (MUSHRA)~\cite{MUSHRA} listening tests at a 48 kHz sample rate. Each of these tests included an unencoded hidden reference, the 3.5~kHz and 7~kHz low-pass filtered versions of the unencoded signals, and one or more coded signals. The codecs used in the training set were Advanced Audio Coding (AAC)~\cite{AAC}, High-Efficiency Advanced Audio Coding (HE-AAC v1 and v2)~\cite{HE-AAC}, and Dolby AC-4~\cite{ac4_IEEE} spanning a wide range of bitrates. We included both speech, music, and a mix of speech and music signals in the training set, but we did not use any listening tests that included a dedicated speech codec. Furthermore, from our audio codec listening experience, we are aware that listeners tend to find it difficult to weigh their preference between spatial coding and waveform coding artifacts when presented concurrently. From our initial experiments, we also observed similar confusion with the predicted quality score from Stereo InSE-NET. Therefore, we enlarged the training data with listening tests that include \emph{hybrid stereo coding}, i.e., codecs whose low bands are coded with waveform preserving stereo coding tools and high bands are coded with parametric stereo coding tools. For this purpose, we included listening tests which included a few experimental codecs that led to the development of the stereo coding tools in the AC-4 standard~\cite{ac-4Part1}. Figure 1 plots the histogram of the mean MUSHRA scores over the entire listening test database, showing a wide range of perceptual quality scores. Note that listening test excerpts can be of different lengths, ranging up to a maximum length of 56.48s in our training set. Therefore, we extended the smaller excerpts to the maximum length of 56.48s by zero-padding them on both sides of the excerpt.

\subsection{Test sets}
\label{subsection:testset}

\begin{figure}[t]
\begin{center}
  \includegraphics[width=1.1\linewidth]{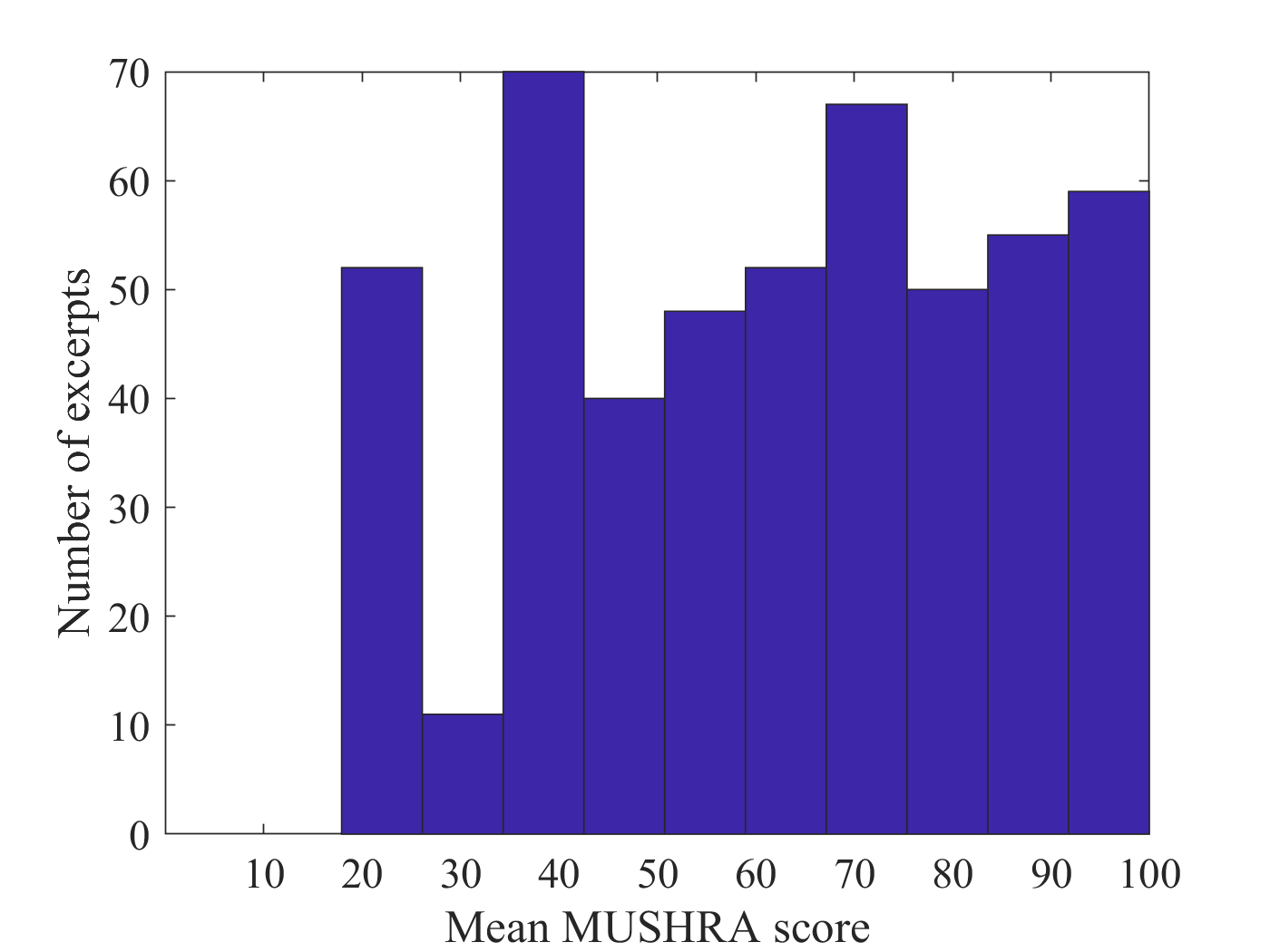} 
  \caption{Histogram of mean MUSHRA scores from the stereo listening test database for testing our model.}
  \label{fig:USAC-VT1and2_mushraDistribution}
  \end{center}
\end{figure}
We benchmarked the prediction accuracy of Stereo InSE-NET against the same stereo listening tests presented in our previous work~\cite{InSE-NET}. We used the Unified Speech and Audio Coding (USAC)~\cite{MPEG-USAC} verification listening tests~\cite{usac_lt},\cite{USAC} to evaluate the performance of the Stereo InSE-NET against subjective listening scores. 

USAC verification listening tests~\cite{usac_lt},\cite{USAC} contain 27 excerpts coded with USAC, HE-AAC, and AMR-WB+ with a wide range of bitrates from 8~kbps mono to 96~kbps stereo. These comprehensive verification tests were designed to provide information on the subjective performance of the USAC codec. It consists of three separate listening tests: mono at low bitrates and stereo at both low and high bitrates. All tests were MUSHRA tests, whose quality scale ranges from 0 to 100, where a higher score implies better quality. The histogram of the mean MUSHRA scores for the two stereo listening tests is shown in Figure 2. For the details of these MUSHRA listening tests, interested readers are referred to~\cite{usac_lt},\cite{USAC}. Note that we included the mono listening test because we also wanted to evaluate the accuracy of our stereo model when trained with only stereo listening tests.

\section{Stereo InSE-NET}
\label{section:StereoInSE-NET}

\subsection{Model architecture}
\label{subsection:model_arch}

As shown in Figure 3, our architecture is built upon our previous architecture, with only minimalistic change. The mono version of the InSE-NET operated on the Gammatone spectrograms of the reference-coded (ref.-cod.) signal pair. Gammatone filters are a popular approximation to the filtering performed by the ear. The Gammatone spectrogram can thus be considered as a more perceptually motivated representation than the traditional spectrogram. The most obvious way to extend the mono InSE-NET to stereo is to simply feed in Gammatone spectrograms of ref.-cod. pairs for left (L) and right (R) channels. However, to condition the model with stereo cues, we feed in additionally ref.-cod. pairs of mid (M) and side (S) channels, where $M = 0.5(L+R)$ and $S = 0.5(L-R)$. Thus, the proposed Stereo InSE-NET model operates on the Gammatone spectrograms of L, R, M, and S signals of reference and coded stereo signals. The Gammatone spectrogram of the audio signal is calculated with a window size of 80\,ms, hop size of 20\,ms, and 32 frequency bands ranging from 50~Hz up to 24~kHz. The four resulting Gammatone spectrograms of reference and coded signals are paired and stacked along the channel dimension, which results in an input size of 8$\times$32$\times$frames (channels$\times$bands$\times$frames) to the network. Since the signals in the training set are equalized to a length of 56.48s, the shape of the input layer is 8$\times$32$\times$2824. The core DNN backbone of Stereo InSE-NET is based on the same modules as the mono InSE-NET: a combination of modified Inception (In) and Squeeze-and-Excitation (SE) blocks followed by fully connected layers. Since the design considers stereo/spatial aspects, we name our model Stereo InSE-NET. Intuitively one would expect mono InSE-NET should have already learned some useful weights for the coded audio quality prediction task. So, we transferred the weights from its first two Inception blocks to the Stereo InSE-NET (except for the input layer dealing with the side-channel) and retrain with listening tests, i.e., transfer learning. The benefits of transfer learning will be shown in Section 3.2.

\begin{figure*}
  \centering
  \includegraphics[width=.74\linewidth]{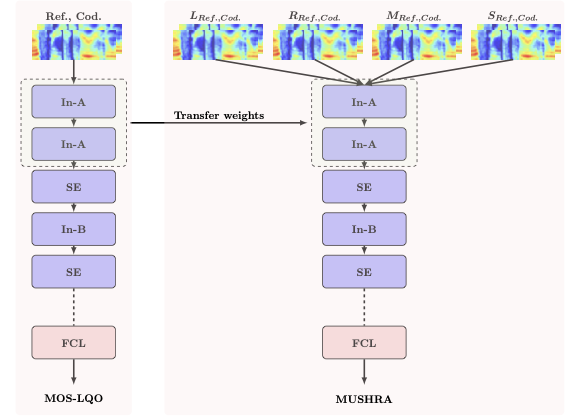} 
  \caption{Block diagrams of the mono InSE-NET~\cite{InSE-NET} (left) and the proposed Stereo InSE-NET model (right). Input to the Stereo InSE-NET is ref.-cod. pairs of Gammatone spectrograms (of L, R, M, and S signals) which are computed in a manner identical to ViSQOL-v3~\cite{ViSQOLgithub}. Except for the input layer dealing with the side-channel, Inception block A, and Inception blocks A and B are initialized from the pre-trained mono InSE-NET.}
  \label{fig:StereoInSE-NET}
\end{figure*}

\begin{table}[t]
    \setlength\tabcolsep{1.8pt}
    \vspace{-0.4cm}
    \caption{Architectures and parameters of the Stereo InSE-NET model. Note that compared to its mono counterpart~\cite{InSE-NET}, only the shape of the input layer has changed.}
    \vspace{-0.4cm}
    \begin{center}
        \scriptsize{
        \begin{tabular}{|l||r|r|r|r|r|}
            \hline
            \multicolumn{1}{|c||}{\textbf{\backslashbox{\textbf{Layer}}{\textbf{Shape}}}}   & \multicolumn{1}{c|}{\textbf{\begin{tabular}[c]{@{}c@{}}Output\end{tabular}}} & \multicolumn{1}{c|}{\textbf{\begin{tabular}[c]{@{}c@{}}Horizontal \\ Conv.\\ (w$\times$h\,/\,ch)\end{tabular}}} & \multicolumn{1}{c|}{\textbf{\begin{tabular}[c]{@{}c@{}}Vertical \\ Conv. \\ (w$\times$h\,/\,ch)\end{tabular}}} & \multicolumn{1}{c|}{\textbf{\begin{tabular}[c]{@{}c@{}}Normal \\ Conv. \\ (w$\times$h\,/\,ch)\end{tabular}}} & \multicolumn{1}{c|}{\textbf{\begin{tabular}[c]{@{}c@{}} Average \\ Pooling  \\ (w$\times$h)  \end{tabular}}} \\ \hline\hline
            \textbf{Input}   & 8$\times$32$\times$2824   &   &   &   &   \\ \hline
            \textbf{Inception A}   & 208$\times$16$\times$180   & 3$\times$7 / 64   & 7$\times$3 / 64   & 1$\times$1 / 64   & 5$\times$5  \\ \hline
            \textbf{Inception A}   & 224$\times$16$\times$90   & 3$\times$7 / 64   & 7$\times$3 / 64   & 1$\times$1 / 64   & 5$\times$5   \\ \hline
            \textbf{SE}   & 224$\times$16$\times$90   &   &   &   &   \\ \hline
            \textbf{Inception B}   & 256$\times$16$\times$45   & 3$\times$5 / 64   & 5$\times$3 / 64   & 1$\times$1 / 64   & 5$\times$5   \\ \hline
            \textbf{SE}   & 256$\times$16$\times$45   &   &   &   &   \\ \hline
            \textbf{Inception C}   & 256$\times$14$\times$22   & 3$\times$3 / 64   & 5$\times$5 / 64   & 1$\times$1 / 64   & 3$\times$3   \\ \hline
            \textbf{SE}   & 256$\times$14$\times$22   &   &   &   &   \\ \hline
            \textbf{\begin{tabular}[c]{@{}l@{}}Adaptive\\ AvgPool\end{tabular}} & 256$\times$4$\times$4   &   &   &   &   \\ \hline
            \textbf{FCL 1}   & 3200$\times$1   &   &   &   &   \\ \hline
            \textbf{FCL 2}   & 512$\times$1   &   &   &   &   \\ \hline
            \textbf{FCL 3}   & 1$\times$1   &   &   &   &   \\ \hline
        \end{tabular}
        }
    \end{center}
    \label{tbl:parameters}
\end{table}

Our proposed Stereo InSE-NET has 15.25M parameters and its architectural and parameter details are tabulated in Table 1. The number of parameters is only $0.022\%$ more than the mono version which is attributed to (six) additional input channels. As discussed for the mono model, the kernel sizes employed in the Inception blocks (A, B, and C) considered several possible kernel shapes and sizes (3$\times$3, 3$\times$5, up to 9$\times$9). A parametric grid search was performed to identify the optimal kernel sizes for each layer. In Table 1, we directly present the optimal kernel sizes of the Inception blocks as obtained for our mono model.  Note that for the stereo model, we have not searched for optimal kernel sizes for each layer. 
\begin{figure}[t]
\begin{center}
  \includegraphics[width=1\linewidth]{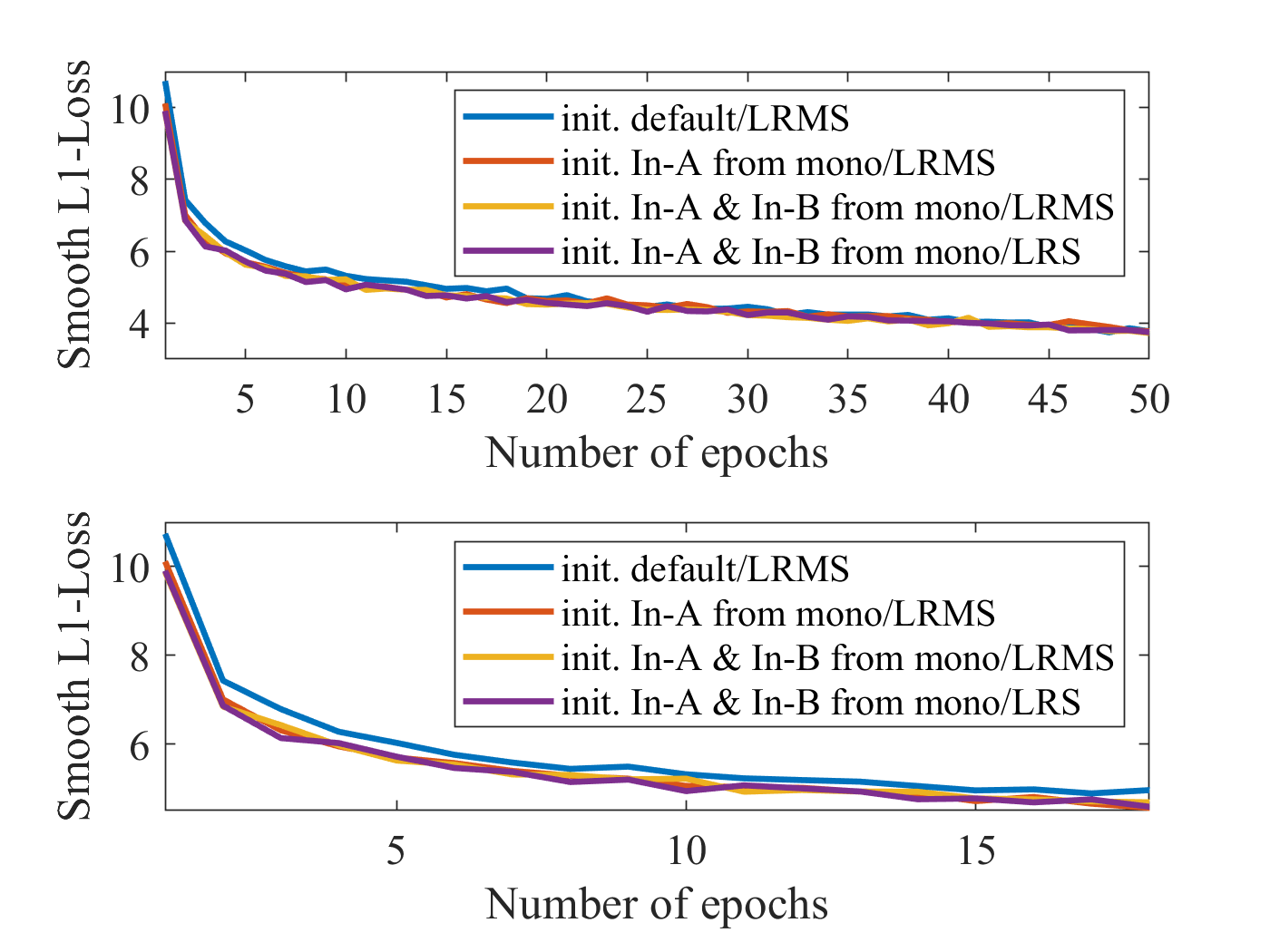} 
  \caption{(Top) Training losses with default PyTorch initializer~\cite{He_2015} (blue); Inception block A (red), and both Inception blocks A and B (yellow) initialized from pre-trained mono InSE-NET. Both blocks are initialized from a pre-trained mono InSE-NET, but without mid-channel (purple). (Bottom) Zoomed-in version of top plot.}
  \label{fig:training_losses}
  \end{center}
\end{figure}

It is interesting to note that in~\cite{dplm} the authors evaluated a variety of different convolutional network building blocks as the feature extraction block, and found Inception resulted in the best features for the spatial audio localization~\cite{dplm},\cite{SAQAM} and quality metric~\cite{SAQAM}. They used a 6-block Inception network operating on phase and magnitude spectrogram, however, unlike ours, they did not do any specific kernel size modification for handling audio signals.

\subsection{Training configuration}
\label{subsection:training_config}

We attempt to leverage the domain knowledge which has been learned during the training of mono InSE-Net with abundant synthetic mono data and transfer these weights of the first two layers of Inception blocks to Stereo InSE-NET rather than training from scratch. There are two advantages of such a strategy: first, diverse, well-structured, and carefully annotated stereo listening tests are still a rare resource, and second, initializing the stereo model with the weights from the mono model would be beneficial for optimization. We then retrained the stereo model with typical MUSHRA listening tests~\cite{MUSHRA}. For a ref.-cod. pair, our training target is the average of the MUSHRA scores rated by listeners. Similar to mono InSE-NET, we made use of the smooth L1-loss~\cite{SmoothL1LossPytorch} for training.  

The training dataset was first normalized and partitioned randomly into $80\%$ for training and $20\%$ for validation. A 5-fold cross-validation is applied to ensure that the model could make full use of limited data. We keep the optimal kernel sizes and learning rates the same as we experimented in the mono InSE-Net and selected a learning rate of $10^{-4}$ and a batch size of 8. Batch normalization and ReLU as activation functions were applied after all convolutional layers. The setup is implemented with PyTorch and was trained for 10 epochs for each fold (i.e., 50 epochs in total) on an Nvidia GTX 1080 Ti GPU with Adam optimizer. The model is trained and evaluated under the following criteria: smooth L1 Loss, mean squared error (MSE), Spearman's correlation coefficient ($R_s$), and Pearson's correlation coefficient ($R_p$).  

The decay of training losses is depicted in Figure 4. We plotted the training losses with the default PyTorch initializer, i.e., Kaiming uniform~\cite{He_2015}. Next, only the Inception block A, and both Inception blocks A and B are initialized from pre-trained mono InSE-NET. Note that we did not initialize the input layer dealing with the side-channel from the pre-trained mono InSE-NET. In the zoomed-in plot, we notice that the initialization from a mono model leads to starting from a lower loss and consequently leads to faster convergence. We also experimented with initializing subsequent blocks from the mono model, but we did not observe any benefit in going beyond initializing Inception block B. Next, we also experimented without mid-channel as the input to the Stereo InSE-NET. We can observe from the plot that for this dataset, it is possibly sufficient to exclude the mid-channel.

\section{Results and Discussion}
\label{section:Results}

We benchmark against ViSQOL because it has been reported in \cite{Fraunhofer}, that out of all objective measures designed to evaluate codecs, ViSQOL shows the best correlation with subjective scores and achieves high and stable performance for all content types. Similarly, it was reported in~\cite{smaq_netflix} that overall VISQOL performed very well across five different datasets (see, Table II in~\cite{smaq_netflix}, overall correlation with ViSQOL as the teacher). Thus, we will benchmark against the latest version of ViSQOL-v3 (operating in audio mode)~\cite{ViSQOLgithub}.

\begin{table}[t]
\setlength\tabcolsep{7pt}
\caption{Performance of ViSQOL-v3, mono InSE-NET, and Stereo InSE-NET on the two stereo listening tests. The table shows the correlation coefficients ($R_p$ and $R_s$) between predicted objective scores and subjective (MUSHRA) scores from the stereo listening tests. The Stereo InSE-NET (a)-(c) indicates training with different data.}
\vspace{-0.4cm}
\begin{center}
\scriptsize{
\begin{tabular}{|l||c|r|c|r|}
\hline
\multirow{2}{*}{\backslashbox{\textbf{Model}}{\textbf{Metric}}}   & \multicolumn{2}{c|}{\textbf{Low Bitrates}}   & \multicolumn{2}{c|}{\textbf{High Bitrates}}      \\ \cline{2-5} 
   & $\mathbf{R_p}$   & \multicolumn{1}{c|}{$\mathbf{R_s}$} & $\mathbf{R_p}$   & \multicolumn{1}{c|}{$\mathbf{R_s}$} \\ \hline\hline
\textbf{ViSQOL-v3}          & \multicolumn{1}{r|}{0.777} & 0.782   & \multicolumn{1}{r|}{0.825} & 0.906   \\ \hline
\textbf{Mono InSE-NET}      & \multicolumn{1}{r|}{0.806} & 0.788   & \multicolumn{1}{r|}{0.847} & 0.895   \\ \hline
\textbf{Stereo InSE-NET (a)}    & \multicolumn{1}{r|}{0.888} & 0.838   & \multicolumn{1}{r|}{0.892} & 0.874   \\ \hline
\textbf{Stereo InSE-NET (b)}    & \multicolumn{1}{r|}{0.897} & 0.861   & \multicolumn{1}{r|}{0.907} & 0.899   \\ \hline
\textbf{Stereo InSE-NET (c)}    & \multicolumn{1}{r|}{0.915} & 0.880   & \multicolumn{1}{r|}{\textbf{0.912}} & \textbf{0.911}   \\ \hline
\textbf{Stereo InSE-NET w/o M (c)}    & \multicolumn{1}{r|}{\textbf{0.922}} & \textbf{0.900}   & \multicolumn{1}{r|}{0.910} & 0.910   \\ \hline
\end{tabular}}
\end{center}
\label{tbl:stereo_results}
\end{table}

We used the Spearman rank-order correlation coefficient ($R_s$) to measure the prediction monotonicity of the models and the Pearson linear correlation coefficient ($R_p$) to measure the prediction linearity. For both $R_p$ and $R_s$, larger values denote better performance. The prediction accuracy of the Stereo InSE-NET on two stereo listening tests is presented in Table 2.  We benchmarked the prediction accuracy of Stereo InSE-NET against the same stereo listening tests presented in our previous work. Internally, ViSQOL downmixes stereo (or any multi-channel signal) to a mono mid-signal and then predicts the MOS. One can observe the prediction accuracy of ViSQOL in the first row. With mono InSE-NET, we compute the mid-signal and feed it into the mono model. Mono InSE-NET results in a slightly better $R_p$ than ViSQOL on the two stereo listening tests as shown in the second row. Moreover, the mono InSE-NET displays a higher accuracy in estimating the quality of excerpts encoded at high bitrates. In the subsequent rows, we demonstrate improving prediction accuracy by training the Stereo InSE-NET with the following data: 

\begin{enumerate}[label=(\alph*)]
\item Stereo listening tests (as described in Section 2.1);
\item Synthetically augmenting (a) by swapping the left- and right-channel of all the audio signals, but with unchanged MUSHRA scores;
\item Additional hybrid stereo coding (as described in Section 2.1) listening tests along with swapping the left- and right-channel of all the audio signals, but keeping MUSHRA scores unchanged.
\end{enumerate}

In the third row in Table 2, we show the impact of training our stereo model with stereo listening tests (data: a). There is a boost in $R_p$. However, there is a slight degradation in $R_s$, suggesting probably the training data was insufficient. In the fourth row, we show the results of training with synthetic listening test data augmentation (data: b). Now both the correlation measures are improved. With the inclusion of hybrid stereo coding listening tests along with listening test data augmentation (data: c), we observe an additional improvement in the fifth row. Overall, we demonstrate a significant improvement ($12\%$ and $6\%$ improvement of Pearson's and Spearman's Rank correlation coefficient, respectively) over ViSQOL-v3 by training the Stereo InSE-NET with both stereo listening tests and synthetically augmented training data. Furthermore, in our ablation experiments, we found that quality prediction accuracy improves with the inclusion of mid- and side-channels. Particularly, the inclusion of the side-channel improves the prediction accuracy towards lower bitrates. In the sixth row, we show that to save computational complexity, it is perhaps sufficient to exclude the mid-channel. We observe a surprising boost in $R_s$ at lower stereo bitrates, but we attribute this result possibly (still) due to the limited amount of training data.

Next, we evaluate the performance of the stereo model with the mono listening test. Note that our stereo model is not trained with mono-listening tests. With our stereo model, we evaluate a dual-mono (i.e., stereo signal with L = R) ref-coded pair.  In \cite{InSE-NET}, we examined the correlation between different objective scores (PEAQ-Advanced \cite{peaqOnline}, ViSQOL-v3 \cite{ViSQOLgithub}, and mono InSE-NET) against subjective scores from the mono USAC verification listening test. In Table 3, we list the results together with the Stereo InSE-NET. We can observe in the last row that even though our model was not trained with mono listening tests, the results display a strong correlation between the prediction of the Stereo InSE-NET and the subjective quality score. These results show, that programmatically generated training data, informed with domain-specific know-how is a powerful technique. But including listening tests as training data still provides a benefit in improving the quality prediction accuracy.

\begin{table}[t]
\setlength\tabcolsep{9pt}
\caption{Performance of PEAQ, ViSQOL-v3, mono InSE-NET, and Stereo InSE-NET with respect to the mono USAC verification listening test. The table shows the correlation coefficients ($R_p$ and $R_s$) between predicted objective scores and subjective (MUSHRA) scores from the mono listening test.}
\vspace{-0.4cm}
\begin{center}
\scriptsize{
\begin{tabular}{|l||r|r|}
\hline
   \backslashbox{\textbf{Model}}{\textbf{Metric}} & \multicolumn{1}{c|}{\textbf{$\mathbf{R_{p}}$}} & \multicolumn{1}{c|}{\textbf{$\mathbf{R_{s}}$}} \\ \hline\hline
\textbf{PEAQ Advanced}          & 0.650 & 0.700\\  \hline
\textbf{ViSQOL-v3}              & 0.810 & 0.840\\  \hline
\textbf{Mono InSE-NET}          & 0.830 & 0.835\\  \hline
\textbf{Stereo InSE-NET}        & \textbf{0.905} & \textbf{0.903}\\  \hline
\end{tabular}}
\end{center}
\label{tbl:mono_results}
\end{table}

Furthermore, we examine the performance of our proposed stereo model along with the mono model on the individual codecs, namely AMR-WB+, HE-AAC, and USAC. The reference and two anchors are included in each examination, and corresponding $R_p$ and $R_s$ are listed in Tables 4a to 4c.

\begin{table}[t]
\setlength\tabcolsep{5pt}
\centering
\caption{Performance of mono and Stereo InSE-NET for various codecs in USAC verification listening tests. The table shows the correlation coefficients ($R_p$ and $R_s$) between predicted objective scores and subjective (MUSHRA) scores from (a) stereo low-bitrate, (b) stereo high-bitrate, and (c) mono listening tests.}

\vspace{-0.4cm}
\begin{center}
\scriptsize{
\begin{tabular}{|l||c|r|c|r|}
\hline
\multirow{2}{*}{\backslashbox{\textbf{Codec}}{\textbf{Metric}}}   & \multicolumn{2}{c|}{\textbf{Mono InSE}}   & \multicolumn{2}{c|}{\textbf{Stereo InSE}}      \\ \cline{2-5} 
   & $\mathbf{R_p}$   & \multicolumn{1}{c|}{$\mathbf{R_s}$} & $\mathbf{R_p}$   & \multicolumn{1}{c|}{$\mathbf{R_s}$} \\ \hline\hline
\textbf{AMR-WB+}    & 0.868   & 0.842   & \textbf{0.960}   & \textbf{0.904}   \\ \hline
\textbf{HE-AAC}     & 0.830   & 0.790   & \textbf{0.945}   & \textbf{0.877}   \\ \hline
\textbf{USAC}       & 0.891   & 0.860   & \textbf{0.976}   & \textbf{0.943}   \\ \hline
\end{tabular}}
\end{center}
\subfloat[\label{tbl:usac2}]
\newline
\vspace{-0.4cm}
\begin{center}
\scriptsize{
\begin{tabular}{|l||c|r|c|r|}
\hline
\multirow{2}{*}{\backslashbox{\textbf{Codec}}{\textbf{Metric}}}   & \multicolumn{2}{c|}{\textbf{Mono InSE}}   & \multicolumn{2}{c|}{\textbf{Stereo InSE}}      \\ \cline{2-5} 
   & $\mathbf{R_p}$   & \multicolumn{1}{c|}{$\mathbf{R_s}$} & $\mathbf{R_p}$   & \multicolumn{1}{c|}{$\mathbf{R_s}$} \\ \hline\hline
\textbf{AMR-WB+}    & 0.864   & 0.852   & \textbf{0.955}   & \textbf{0.925}   \\ \hline
\textbf{HE-AAC}     & 0.871   & 0.925   & \textbf{0.946}   & \textbf{0.949}   \\ \hline
\textbf{USAC}       & 0.909   & 0.920   & \textbf{0.964}   & \textbf{0.942}   \\ \hline
\end{tabular}}
\end{center}
\subfloat[\label{tbl:usac3}]
\newline
\vspace{-0.4cm}
\begin{center}
\scriptsize{
\begin{tabular}{|l||c|r|c|r|}
\hline
\multirow{2}{*}{\backslashbox{\textbf{Codec}}{\textbf{Metric}}}   & \multicolumn{2}{c|}{\textbf{Mono InSE}}   & \multicolumn{2}{c|}{\textbf{Stereo InSE}}      \\ \cline{2-5} 
   & $\mathbf{R_p}$   & \multicolumn{1}{c|}{$\mathbf{R_s}$} & $\mathbf{R_p}$   & \multicolumn{1}{c|}{$\mathbf{R_s}$} \\ \hline\hline
\textbf{AMR-WB+}    & 0.889   & 0.856   & \textbf{0.948}   & \textbf{0.922}   \\ \hline
\textbf{HE-AAC}     & 0.853   & 0.791   & \textbf{0.945}   & \textbf{0.887}   \\ \hline
\textbf{USAC}       & 0.873   & 0.881   & \textbf{0.950}   & \textbf{0.939}   \\ \hline
\end{tabular}}
\end{center}
\subfloat[\label{tbl:mono}]

\end{table}

The proposed stereo model results in a significantly better $R_p$ and $R_s$ than the mono model. Even though we have not trained our model with AMR-WB+ (a speech codec-based system) and USAC codec (which uses a dedicated speech codec for speech signals at low bitrates), the overall performance of our model on these unseen codecs has significantly improved. The proposed stereo model has performed homogeneously well on the experimented mono and stereo codecs. The estimation of coding quality (particularly $R_s$) of HE-AAC, under this comparison, is unexpectedly the worst among the three experimented codecs, even though our model is trained on the listening test excerpts encoded with HE-AAC and AAC. One possible explanation for this result is one of the lowest bitrates in the low bitrate mono and stereo listening tests is unseen during the training.

\section{Conclusion and Discussion}
\label{section:ConclusionDiscussion}

In this paper, we present Stereo InSE-NET: a novel DNN-based coded stereo audio quality prediction model. The model is an extension of our previous mono InSE-NET. The core DNN backbone of Stereo InSE-NET is based on the same modules as the mono InSE-NET: a combination of modified Inception and Squeeze-and-Excitation blocks followed by fully connected layers.  The new design incorporates inter-channel aspects in the design by conditioning the DNN-backbone with Gammatone spectrograms of left, right, mid, and side channels. We demonstrate that weights learned from the mono quality prediction model fine-tuned for the stereo quality prediction task leads to faster convergence. Furthermore, we show that utilizing both real listening test data, and synthetically derived data from listening tests leads to an improvement in prediction accuracy.

In the future, we would like to extend our model to handle both coded stereo and binaural audio quality prediction. Assuming spatial audio content of more than two channels can be rendered to a binaural representation~\cite{peaq-mc},\cite{seo2013perceptual} we would then also investigate our model with multichannel subjective listening test data. 

Finally, our model enables faster prediction.  An un-optimized PyTorch implementation of our model (excluding the Gammatone spectrogram computation frontend) runs at 63.8x real-time on a CPU with the latest PyTorch version 1.12.1. Including the Gammatone spectrogram computation frontend in Python, our complete Stereo InSE-NET model runs at 1.2x real-time on a CPU, which is slightly faster than ViSQOL-v3 (which runs at 1.1x real-time on CPU). However, it is important to reiterate that ViSQOL-v3 consists of traditional signal processing-based blocks which are fully implemented in C++, whereas our proposed Stereo InSE-NET model is a unoptimized Python and PyTorch implementation of a deep CNN-based model. Furthermore, ViSQOL-v3 computes a pair of Gammatone spectrograms, whereas we compute four pairs of Gammatone spectrograms. Therefore, even the slight advantage in computational efficiency over ViSQOL-v3 is a positive outcome.

\bibliographystyle{IEEEtran}

\end{document}